\documentclass[prb,preprint]{revtex4-1} 
\usepackage[utf8]{inputenc}
\usepackage{amsmath, amssymb}
\usepackage[margin=1in]{geometry}
\usepackage{graphicx}
\usepackage{hyperref}
\usepackage{tikz}
\usepackage{todonotes}
\usepackage{caption}
\usepackage{subcaption}

\begin{document}

\title{$N$-body linear force law allowing analytic solutions }
\author{Joseph West}
\email{joseph.west@indstate.edu}
\author{Sean P. Bartz}
\email{sean.bartz@indstate.edu}
\affiliation{Dept.~of Chemistry and Physics, Indiana State University, Terre Haute, IN 47809}
\date{May 1,~2025}
\begin{abstract}
    
We present a pair-wise force law in a system of $N$ particles that produces analytic solutions for an arbitrary number of particles, masses, and initial conditions. 
Each pair of particles interacts via a force that is proportional to the product of their masses and their separation distance, with the force directed radially.
We show that, despite the $N$-body interaction, each particle behaves as if it interacts only with the center of mass of the system.
This effective one-body interaction behaves as Hooke's Law with a common frequency for all particles, with the familiar analytic solutions for the trajectories.
With these analytic solutions, it is possible to efficiently simulate a collection of these particles and incorporate other external forces. As an example, we simulate the particles within an adiabatically expanding container and calculate pressure and temperature in both the attractive and repulsive cases.

\end{abstract}
\maketitle

\section{Introduction}
The general $N$-body problem has long been a challenge in analytical mechanics. Newtonian gravity has been of particular interest, where the three-body problem admits analytical solutions only in special cases. \cite{Sundman1913,Moore1993, Chenciner2000, Li2018, Przybylska2023,uvakov2014,Delshams2018, Wild1980, Harmon2003, Khandekar1972}
Even with numerical solutions, the chaotic nature of $N$-body problems means that arbitrary systems are unstable, presenting pedagogical challenges for students interested in simulating such a system.

This paper presents an investigation of a pairwise force law with analytic solutions in an $N$-body system. Intriguingly, this force law is shown to be equivalent to a central force acting on each particle, and each individual particle exhibits motion independent of its immediate neighbors. 
 Given their mathematical simplicity, these solutions can be conveniently integrated into undergraduate courses on advanced mechanics, within units on harmonic oscillators or gravitational interactions.  
Moreover, due to their computational efficiency, these systems can be accurately simulated with large $N$ values on personal computers, thus rendering them suitable for inclusion in computational projects or courses on numerical methods. 

 The absence of short range interactions is convenient for quantum systems based on similar interactions, as the Hamiltonian is separable \cite{Calogero1969, Calogero1971,  Lim1971, Vladimiroff1974,BialynickiBirula1985,Crandall1984}.  
 However, this simplicity means that the classical system in isolation lacks the interesting orbital dynamics of Newtonian gravity. 
 For this reason, further applications of this force law are most intriguing when paired with an external interaction. In this paper, we explore a thermodynamic application with the system in a container and suggest follow-up student projects.

The paper is organized as follows.  In Section \ref{sec:force} the force law is introduced and the equivalence of an $N$-body interaction to a one-body interaction with the center of mass is shown.  In Section \ref{sec:applications}, we present thermodynamic results for attractive and repulsive systems as a function of volume.  Conclusions are provided in Section \ref{sec:conclusion}. 
The collision algorithm is presented in Appendix \ref{sec:collision_algorithm}, while ideas for possible research projects for advanced undergraduate students involving these two systems are discussed in Appendix \ref{sec:projects}. 
Supplemental materials include a number of short python programs, including one for plotting the orbits of an N-particle system given a specified set of masses and initial conditions based on the analytic trajectories. A similar program that runs a numerical simulation of a similar system directly from the pair-wise forces is also included.

\section{Force Law and General Solution} \label{sec:force}
Bertrand’s theorem shows that the only central forces that guarantee closed orbits are the inverse square and linear force laws. \cite{bertrand1873} Inspired by this, we introduce a toy model of a linear gravity-like force 
\begin{equation}
\mathbf{F} = - Jm_1 m_2 \mathbf{r}, \label{eq:force_def}
\end{equation}
where $ J $ is a constant with units of  N/kg$^2$m and can be positive (attractive force) or negative (repulsive). The force is that on mass $m_2$ due to mass $m_1$, and $\mathbf{r}$ is the vector from $m_1$ to $m_2$.  The force is gravity-like in that the force between particles is proportional to the mass of both particles.  It is spring-like in that it is linear.  In the attractive case, the force is equivalent to a system of masses connected pair-wise by springs with spring constants  
\begin{equation}
    k_{ij} = Jm_im_j 
\end{equation}
for each pair, as illustrated in Fig. \ref{fig:force_representation}. 
 
\begin{figure}
\begin{minipage}{0.5\textwidth}
\begin{tikzpicture}[scale=.75] 
  \def\scalefactor{0.02} 
  \def\relativesize{1} 
  \def\positionscale{1} 
  
  \tikzset{label style/.style={font=\fontsize{14}{14}\selectfont}}
  
  \node[draw, circle, fill=blue, minimum size=0.5*\relativesize cm] (m1) at (0*\positionscale,-1*\positionscale) {};
  \node[draw, circle, fill=blue, minimum size=1*\relativesize cm] (m2) at (5*\positionscale,0*\positionscale) {};
  \node[draw, circle, fill=blue, minimum size=2*\relativesize cm] (m3) at (4*\positionscale,3*\positionscale) {};
  \node[draw, circle, fill=blue, minimum size=3*\relativesize cm] (m4) at (-1*\positionscale,4*\positionscale) {};
  
  \node[left, label style] at (m1.west) {$m_1$};
  \node[right, label style] at (m2.east) {$m_2$};
  \node[above right, label style] at (m3.north) {$m_3$};
  \node[above left, label style] at (m4.north) {$m_4$};
  
  \draw[dashed, line width=0.5*\relativesize*\scalefactor cm] (m1) -- (m2) node[midway, below, label style] {$k_{12}$};
  \draw[dashed, line width=1.0*\relativesize*\scalefactor cm] (m1) -- (m3) node[midway, left, label style] {$k_{13}$};
  \draw[dashed, line width=1.5*\relativesize*\scalefactor cm] (m1) -- (m4) node[midway, left, label style] {$k_{14}$};
  \draw[dashed, line width=2.0*\relativesize*\scalefactor cm] (m2) -- (m3) node[midway, right, label style] {$k_{23}$};
  \draw[dashed, line width=3.0*\relativesize*\scalefactor cm] (m2) -- (m4) node[near end, left, label style] {$k_{24}$};
  \draw[dashed, line width=6.0*\relativesize*\scalefactor cm] (m3) -- (m4) node[midway, above, label style] {$k_{34}$};
\end{tikzpicture}
\end{minipage}%
\begin{minipage}{0.5\textwidth}
\begin{tikzpicture}[scale=.75]
  \def\scalefactor{0.05}
  \def\relativesize{1}
  \def\positionscale{1}
  
  \tikzset{label style/.style={font=\fontsize{14}{14}\selectfont}}
  
  \node[draw, circle, fill=blue, minimum size=0.5*\relativesize cm] (m1) at (0*\positionscale,-1*\positionscale) {};
  \node[draw, circle, fill=blue, minimum size=1*\relativesize cm] (m2) at (5*\positionscale,0*\positionscale) {};
  \node[draw, circle, fill=blue, minimum size=2*\relativesize cm] (m3) at (4*\positionscale,3*\positionscale) {};
  \node[draw, circle, fill=blue, minimum size=3*\relativesize cm] (m4) at (-1*\positionscale,4*\positionscale) {};

  \pgfmathsetmacro{\xcm}{(0*0.25 + 5*0.5 + 4*1 + -1*1.5) / (0.25 + 0.5 + 1 + 1.5)}
  \pgfmathsetmacro{\ycm}{(-1*0.25 + 0*0.5 + 3*1 + 4*1.5) / (0.25 + 0.5 + 1 + 1.5)}

  \node[draw, circle, minimum size=0.05*\relativesize cm] (cm) at (\xcm*\positionscale,\ycm*\positionscale) {};
  
  \draw[dashed, line width=\scalefactor*0.5*\relativesize cm] (m1) -- (cm) node[midway, left, label style] {$k_1$};
  \draw[dashed, line width=\scalefactor*1*\relativesize cm] (m2) -- (cm) node[near start, left, label style] {$k_2$};
  \draw[dashed, line width=\scalefactor*2*\relativesize cm] (m3) -- (cm) node[midway, above, label style] {$k_3$};
  \draw[dashed, line width=\scalefactor*3*\relativesize cm] (m4) -- (cm) node[near start, above right, label style] {$k_4$};
  
  \node[left, label style] at (m1.west) {$m_1$};
  \node[above right, label style] at (m2.east) {$m_2$};
  \node[above right, label style] at (m3.north) {$m_3$};
  \node[above left, label style] at (m4.north) {$m_4$};
\end{tikzpicture}
\end{minipage}

    \caption{Left: a visualization of the $N$-body force interaction. The magnitudes of the effective spring constants $k_{ij}=Jm_im_j$ are represented by the thickness of the lines. Right: a visualization of this $N$-body force reduced to individual forces interacting with the center of mass (shown as an empty circle). The new effective spring constants are $k_i = JMm_i$, where $M$ is the sum of all masses.  }
    \label{fig:force_representation}
\end{figure}
 
\subsection{Solutions for point particles}

We solve the $N$-body problem by showing that it is equivalent to $N$ one-body problems.
Consider a collection of \( N \) particles with total mass \( M \). The equation of motion of particle \( i \) is 
\begin{equation}
    m_i \ddot{\mathbf{r}}_i = -J \sum_{k \neq i} m_i m_k (\mathbf{r}_i - \mathbf{r}_k).
\end{equation}
The contribution from \( k = i \) is zero, so it may be included for convenience. Including this term, the sum becomes
\begin{eqnarray}
   m_i \ddot{\mathbf{r}}_i &=& -J m_i \sum_k m_k (\mathbf{r}_i - \mathbf{r}_k)\\
    &=& -J m_i\left( \mathbf{r}_i \sum_k m_k  -  \sum_k m_k  \mathbf{r}_k \right)
\end{eqnarray}
The sum in the second term is recognizable from the definition of the center of mass, and the equation of motion becomes
\begin{equation}
    m_i \ddot{\mathbf{r}}_i = -JMm_i(\mathbf{r}_i-\mathbf{R}_\mathrm{cm}),
\end{equation}
where $M=\sum_k m_k$ is the total mass of the system and $\mathbf{R}_\mathrm{cm}$ is the position of the center of mass, which need not be stationary. Thus, the particle behaves as if it is governed by Hooke's Law centered at $\mathbf{R}_\mathrm{cm}$, with effective spring constant $k_i=JMm_i, $ as illustrated in Fig. \ref{fig:force_representation}.  

The solutions are the harmonic oscillator solutions 
\begin{equation}
    \mathbf{r}_i = \mathbf{A}\cos(\omega t) +\mathbf{B}\sin(\omega t) +\mathbf{R}_\mathrm{cm}, \label{eq:attractive_trajectory}
\end{equation}
where $\omega=\sqrt{JM}$ for all particles and the constant vectors $\mathbf{A}, \, \mathbf{B}$ are determined as usual from initial conditions. In the absence of external forces, the center of mass of the system moves at constant velocity, and we can use coordinates with $\mathbf{R}_\mathrm{cm}=0$ so that the particles move in closed elliptical orbits with a common frequency, consistent with Bertrand's Theorem. \cite{bertrand1873, taylor2005classical,goldstein2001classical,grossman1996sheer}

Thus, the $N$-body interaction has an analytical solution where the motion of each particle is effectively described as an interaction with the center of mass of the system.
This differs from the general $N$-body problem, which cannot be solved analytically. Additionally, particles in this system remain bound, in contrast to the inverse-square law system, which can eject particles from the system. 



 \subsection{Particle interacting with continuous mass distribution} \label{sec:continuous}
We now generalize the above results to a particle interacting with a continuous mass distribution.  Consider a point mass m at the origin that interacts with a mass $M$ distributed in space within a  volume $V$ that has a position-dependent density $\rho(\mathbf{r})$.  
In that case, the force on \( m \) due to a small volume element \( dV \) within the extended body is
\begin{equation}
d\mathbf{F} = -Jm \left( \rho(\mathbf{r)} dV \right) \mathbf{r}. 
\end{equation} 
The total force is found by integrating over the volume of the extended body
\begin{equation}
\mathbf{F}  = -Jm\int \rho(\mathbf{r}) dV \mathbf{r} = -JmM \mathbf{R}_\mathrm{cm}. \label{eq:extended}
\end{equation}
Notice, due to the linear nature of the force law, there is no concern of singularities in allowing the particle of interest to be located within the larger mass distribution.  From (\ref{eq:extended}), it is clear that the shape, size, and orientation of the mass distribution have no effect on the motion of test mass $m$.  The distributed mass does not need to be a rigid body, and it may rotate in any arbitrary manner, as long as it is rotating about its center of mass, which is how unconstrained bodies rotate. \cite{fowles1999analytical, Thornton_Stephen_T_2022-01-03, Hamill2022} 
 \subsection{Potential energy}
The potential energy \( U \) of the system is also found analytically. The potential energy of a system of particles attached via the properly chosen springs is
\begin{equation}
U = \frac{1}{2} \left[ \frac{1}{2}J\sum_{i,k} m_i m_k (\mathbf{r}_i - \mathbf{r}_k) \cdot(\mathbf{r}_i - \mathbf{r}_k)\right],
\end{equation} 
where the additional factor of \( \frac{1}{2} \) corrects for double counting. The case \( i = k \) does not contribute to the sum, so it need not be removed. Expanding the product yields
\begin{equation}
U = \frac{J}{4} \left[ \sum_{i,k} m_i m_k (r_i^2 + r_k^2 - 2\mathbf{r}_i \cdot \mathbf{r}_k) \right], 
\end{equation}
and separating the terms yields 
\begin{align}
U = \frac{J}{4} \left[ \sum_i m_i\sum_k m_k r_k^2 + \sum_k m_k\sum_i m_i r_i^2 - 2\sum_i m_i \mathbf{r}_i \cdot \sum_k m_k \mathbf{r}_k \right]. \label{eq:potential_energy}
\end{align}
The last term is proportional to the squared magnitude of the location of the center of mass, which is set to be zero.  The summations over mass alone in the first two terms gives the total mass $M$, so the first two terms are identical. The resulting potential energy 
\begin{equation}
U = \sum_k \frac{1}{2} JMm_k r_k^2 \label{eq:potential_energy_simplified}
\end{equation}
is identical to $k$ independent harmonic oscillators with effective spring constant $JMm_k$.

Alternatively, we calculate the work to move the $N$ particles from the origin to their positions $\mathbf{r}_k$ from the force law (\ref{eq:force_def}) directly. The work done in moving a single particle is 
\begin{equation}
W_k  = \int_0^{r_k} JMm_k r_k dr_k,
\end{equation}
as the work is non-zero only for the radial component of displacement. Performing the integration and summing over all particles, 
\begin{equation}
\sum W_k = \sum_k \frac{1}{2} JMm_k r_k^2,
\end{equation}
 which is equal to the potential energy calculation (\ref{eq:potential_energy_simplified}).

\subsection{Repulsive Case}
 
For a repulsive interaction, the above analysis is repeated with $J<0$. The solutions become exponentials
\begin{equation}
    \mathbf{r}_i= \mathbf{C}e^{\Omega t} +\mathbf{D}e^{-\Omega t} +\mathbf{R}_\mathrm{cm} \label{eq:repulsive_trajectory}
\end{equation}
where \( \Omega \equiv \sqrt{-JM} \). The energy analysis is the same as in Section \ref{sec:continuous}, with $J\rightarrow -J$.
Unless $\mathbf{C}=0$, the particle escapes to infinity  with exponentially increasing velocity.
The potential energy of a particle is always negative 
\begin{equation}
    U_k=-JMm_kr_k^2,
\end{equation}
which explains why the kinetic energy grows without bound as the particle escapes to infinity.
 

\section{Applications}\label{sec:applications}
 The existence of analytic solutions allows for efficient simulation of large numbers of particles, effectively modeling a gas of strongly interacting particles. In this section, we use analytic simulations to study thermodynamic properties of such a large collection.

Numerical techniques explicitly evolve the system forward in time given the current state of positions and velocities. It is pedagogically useful to show that this numerical technique produces results that match the analytic solutions of Section \ref{sec:force}. Animating the trajectories using VPython is also instructive, as it effectively illustrates the “non-interaction” of nearby particles. This also demonstrates the benefits of the Euler-Cromer method, which conserves energy in oscillatory motion, while the Euler method does not. 
 Numerical techniques are also needed if the system is subjected to external forces. Examples of these applications are described in Appendix \ref{sec:projects}.

Any numerical technique using the original pair-wise force law \eqref{eq:force_def} must calculate the distance between each particle pair, a problem that scales as $\mathcal{O}(N^2)$. Additionally, errors in the differential equation solver depend on the time step, with the Euler-Cromer method having errors $\mathcal{O}(dt)$. Thus, simulating a large number of particles interacting for a long time is computationally intensive for a student project. This motivates the use of the analytic solutions when possible.

In the case of an isolated system, the analytic solutions \eqref{eq:attractive_trajectory} and \eqref{eq:repulsive_trajectory} give the position and velocity of any particle at arbitrary time. In the case of a system inside a container, a continuous collision algorithm is used to handle any particles that collide with the container during a given time step.  Any  particle colliding with the wall  affects the rest of the masses, because the change in velocity of that single particle affects the center of mass.
The effects of the collisions on the velocity of the center of mass are accumulated while the check for collisions is made. The analytic solutions are updated to account for the change in center of mass velocity. The effects of an adiabatically expanding container are also incorporated. This collision algorithm is elaborated in Appendix \ref{sec:collision_algorithm}.

\subsection{Thermodynamic relationships}

\begin{figure}[htb]
    \centering
    \begin{subfigure}{0.49\textwidth}
        \centering
        \includegraphics[width=\textwidth]{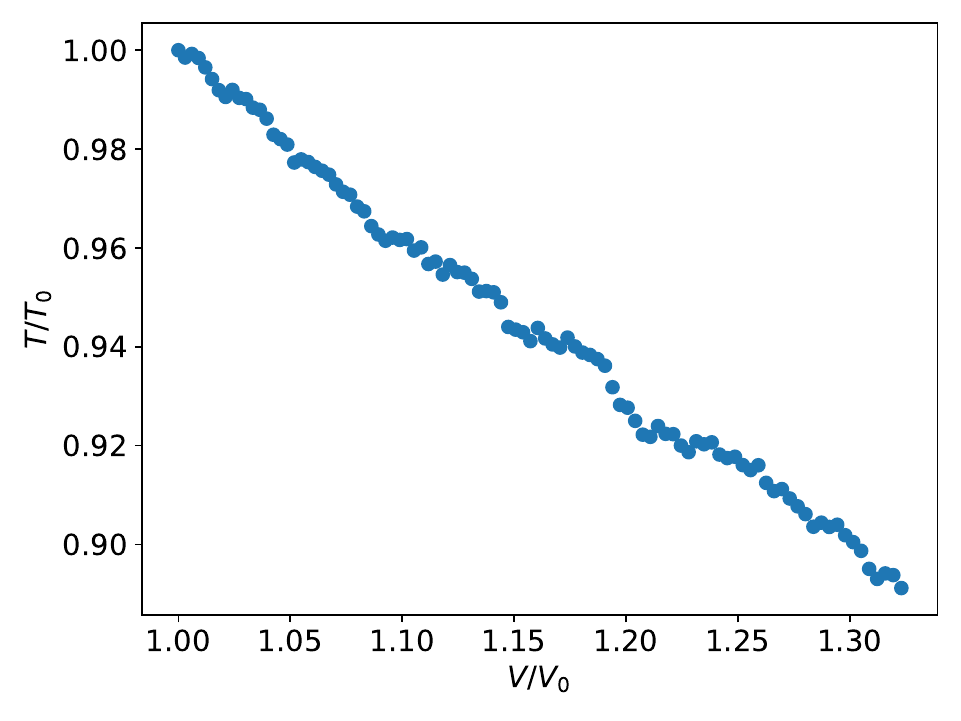}
        \caption{}
    \end{subfigure}
    \hfill
    \begin{subfigure}{0.49\textwidth}
        \centering
        \includegraphics[width=\textwidth]{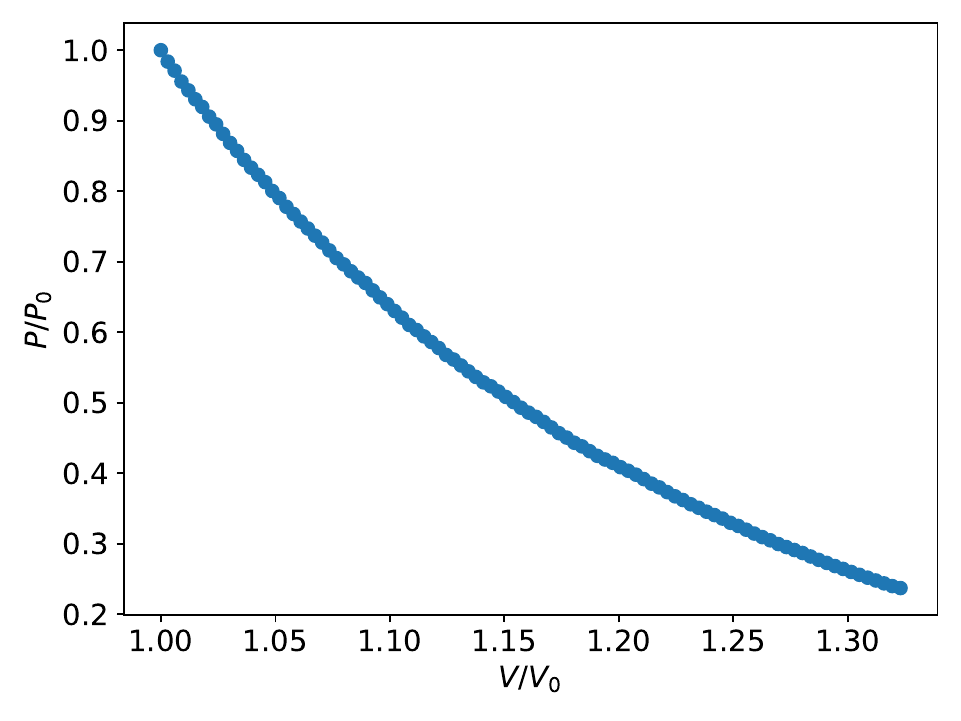}
        \caption{}
    \end{subfigure}

    \begin{subfigure}{0.49\textwidth}
        \centering
        \includegraphics[width=\textwidth]{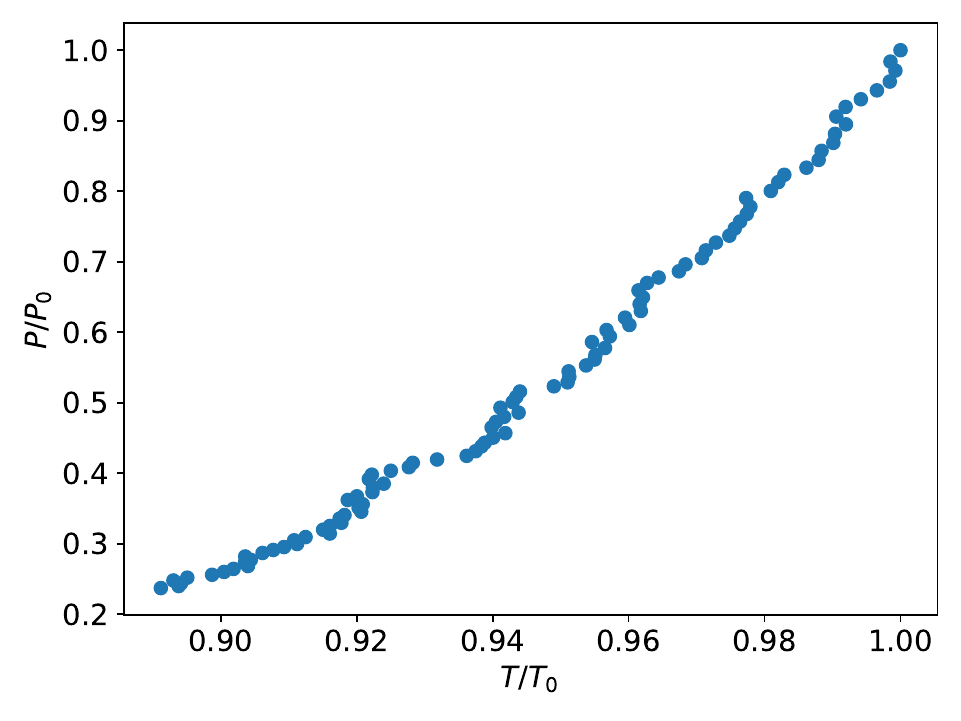}
        \caption{}
    \end{subfigure}
    
    \caption{This figure shows relationships between thermodynamic quantities for the attractive case. Temperature (a) and pressure (b) both decrease as volume is increased. Plot (c) shows the resulting positive correlation between pressure and temperature. From visual inspection, one might expect a power-law relationship among these quantities, but analysis of log-log plots indicates no such behavior. \label{fig:swarm_plots}}
\end{figure}
 
Here the attractive and repulsive systems are examined in a 3D spherical container.  A spherical container is necessary because a rectangular prism will not thermalize the motion of the particles, as the motion along each of the Cartesian coordinates is independent. A collection of 4,000 particles is created with randomized initial positions and velocities within the initial spherical container. The masses are chosen at random to be between 4 and 5 base units.

In contrast to the ideal gas model, systems with long-range interactions do not have well-defined thermodynamic quantities that apply to the entire system. \cite{BerberanSantos1997, Lente2020,MenesesJoo2022} Instead, we examine the pressure and temperature at the boundary of the container.
The pressure is calculated by summing the change in momentum of the particles that hit the wall in each time step,
\begin{equation}
P = \frac{\sum |\Delta p_r|}{4\pi R^2 \Delta t},
\end{equation}
where $p_r$ is the radial component of the particle's momentum R, is the radius of the container, and $\Delta t$ is the computational time step.  
The temperature is determined by averaging the kinetic energy of the particles that hit the wall in a given time interval.  
The pressure and temperature are therefore only defined in terms of the  interaction with the environment via the walls of the container. This is similar to the definition of thermodynamic parameters for bound gravitational systems. \cite{BerberanSantos1997,Lente2020,MenesesJoo2022} In the attractive case, pressure and temperature are well defined only when the container is sufficiently small that the particles interact with the walls of the container.
 
The only quantity held fixed in these simulations is the number of particles. The discontinuous motion of the container and the elastic collisions ensures no work is done on the system, so total energy remains constant.

\begin{figure}[htb]
    \centering
    \begin{subfigure}{0.49\textwidth}
        \centering
        \includegraphics[width=\textwidth]{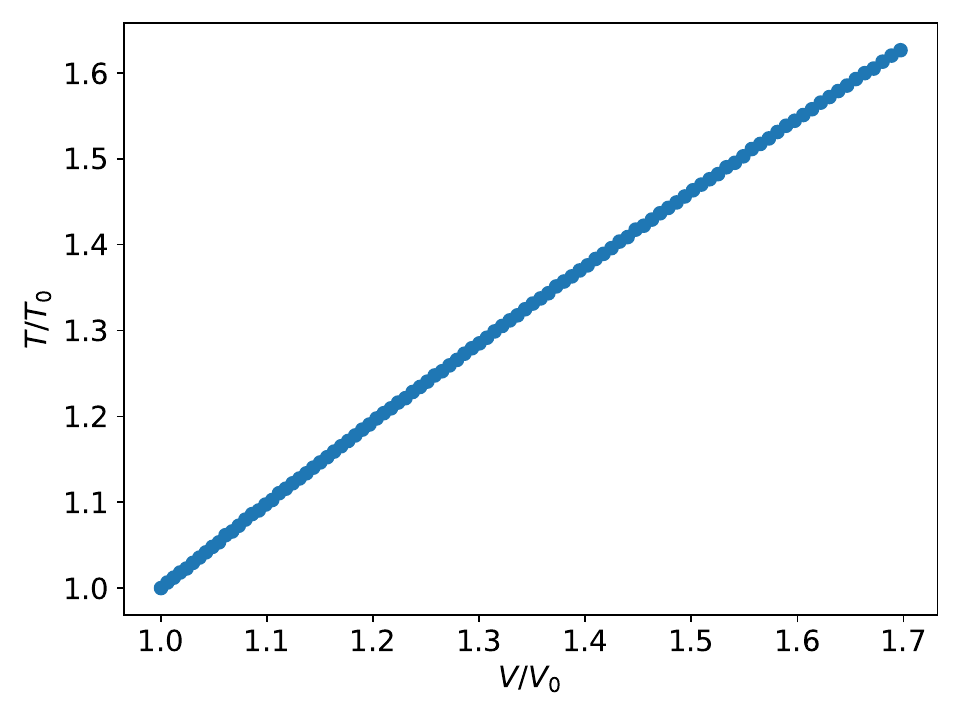}
        \caption{\label{subfig:fireballTV}}
    \end{subfigure}
    \hfill
    \begin{subfigure}{0.49\textwidth}
        \centering
        \includegraphics[width=\textwidth]{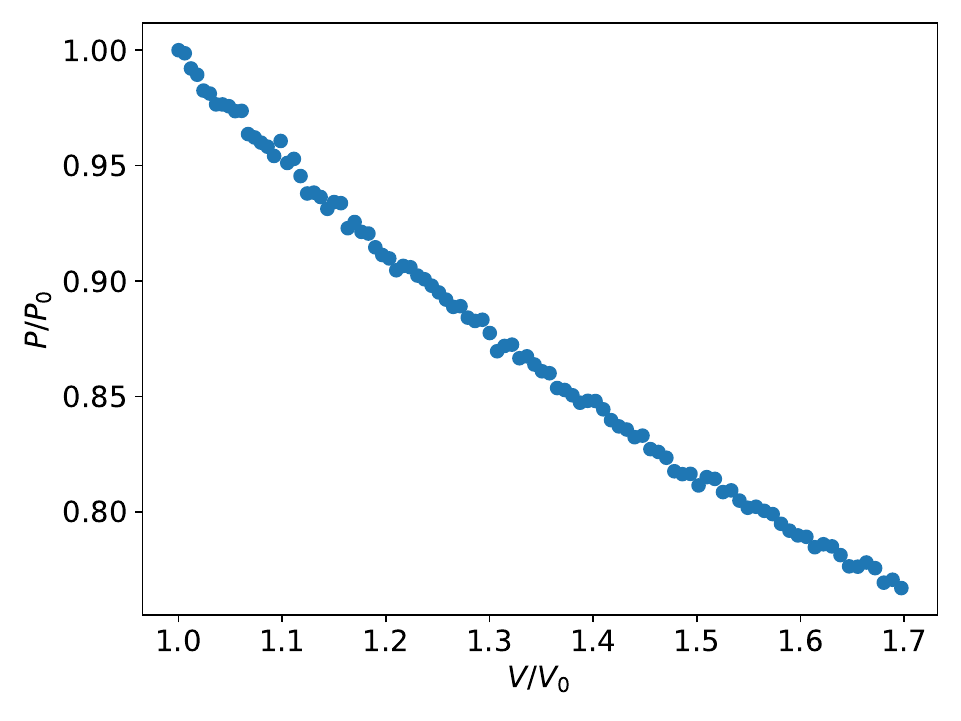}
         \caption{\label{subfig:fireballPV}}
    \end{subfigure}

    \begin{subfigure}{0.49\textwidth}
        \centering
        \includegraphics[width=\textwidth]{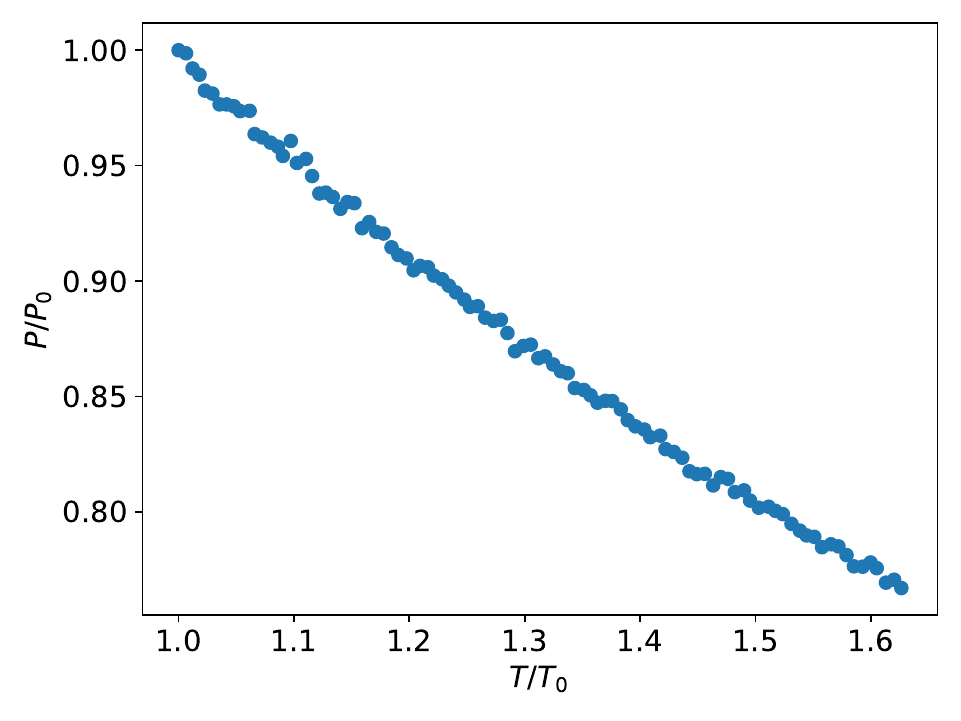}
        \caption{\label{subfig:fireballPT}}
    \end{subfigure}
    
    \caption{Relationships between thermodynamic quantities for the repulsive case. (a) Because the kinetic energy increases without bound, temperature increases with volume. (b) Pressure is inversely proportional to the square root of volume. (c) In contrast to the attractive case, pressure and temperature are negatively correlated. The relationship between temperature and the other quantities cannot be described by a power law.  \label{fig:Fireball}}
\end{figure}

In examining Figs. \ref{fig:swarm_plots}-\ref{fig:Fireball}, it is evident that there is more scatter in the temperature values than in the pressure values. This is because the pressure depends on the radial component of the velocity of the particles that hit the wall, which will equilibrate due to collisions with the spherical container. The temperature also depends on the tangential component of the velocity, which is unaffected by contact with the wall.

For the attractive case, 
 the relationships among pressure, temperature, and volume share some similarity with those for the ideal gas model (see Fig. \ref{fig:swarm_plots}). 
 However, log-log plots indicate the relationships do not follow power laws.

For the repulsive case, an increase in the volume allows the potential energy to grow very large and negative, so that the kinetic energy of the particles impacting the wall of the container increases.  As such, Fig. \ref{fig:Fireball} shows that the temperature increases with increasing volume, contrary to the behavior of attractive systems and ideal gases. However, pressure decreases as a result of the increasing surface area, and we find $P\sim V^{1/2}$ from a log-log fit of the data. The relationship between pressure and volume appears linear on visual inspection of Fig. \ref{subfig:fireballPV}, but a log-log fit reveals that the relationships involving temperature do not follow power laws.

\subsection{Thermodynamic equilibration}

In the simulation, the container is expanded adiabatically in a series of discontinuous jumps, and the particles are allowed some time for thermodynamic equilibration after each expansion. The container radius is increased discontinuously with a time interval $\tau$ between increases, allowing the particles to undergo free expansion into the increased volume.  
Thermalization is allowed for a duration $0.8\tau$, and the thermodynamic quantities are averaged over the final $0.2\tau$ before the subsequent expansion.
The thermalization time is particularly long for the repulsive case, as expanding the container changes the kinetic energy of the system. The values presented in Figs. \ref{fig:swarm_plots} and \ref{fig:Fireball} are the values at the end of the time segment used to equilibrate.

\begin{figure}[htb]
    \centering
    \begin{subfigure}{0.49\textwidth}
\includegraphics[width=\textwidth]{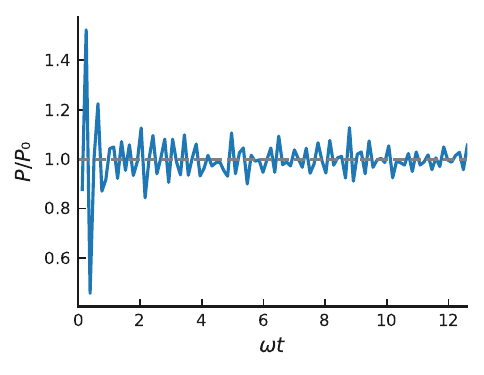}
    \end{subfigure}
    \begin{subfigure}{0.49\textwidth}
        \includegraphics[width=\textwidth]{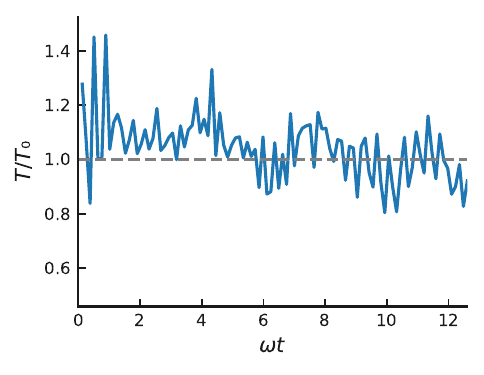}
    \end{subfigure}
   
    \caption{Thermodynamic equilibration over a short time period in a fixed volume is illustrated by the pressure (left) and temperature (right). The values are normalized by the average of the data for $\omega t>6$.}
    \label{fig:equilibration_longer}
\end{figure}

In Fig. \ref{fig:equilibration_longer}, we show a simulation of the attractive system in a fixed container over a short period of time $\omega t=10/\pi$, plotting the values at each time step, with no averaging.  The particles start out in a random configuration, but after interactions with the container walls, the pressure and temperature settle down toward equilibrium values. 


\begin{figure}[htb]
    \centering
    \begin{subfigure}{0.49\textwidth}
        \includegraphics[width=\textwidth]{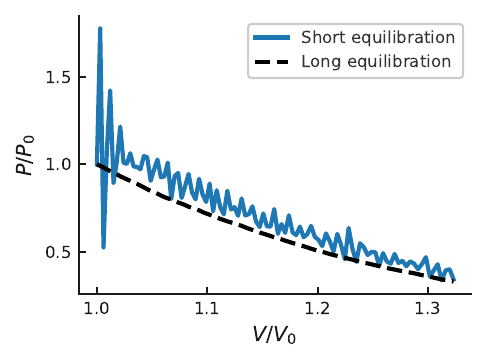}
    \end{subfigure}
    \begin{subfigure}{0.49\textwidth}
        \includegraphics[width=\textwidth]{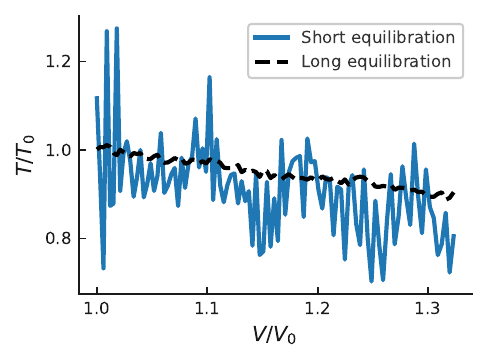}
    \end{subfigure}
    \caption{A comparison of results with different thermal equilibration times. The solid blue lines represent the same results shown in Fig. \ref{fig:swarm_plots}. The orange dashed lines represent simulations where the container is expanded in the same small increments, but with less time for the particles to reach thermal equilibrium. The overall trend is still visible, but the relationships are less clear.}
    \label{fig:equilibration_time_compare}
\end{figure}

We compare the effects of equilibration time in Fig. \ref{fig:equilibration_time_compare}. In the short equilibration simulation, the equilibration time is $\sim 12.7$ times the oscillation period of the particles. The long equilibration is 100 times longer.  When there is less equilibration time, the particles may not settle to the equilibrium state, and there is more noise in the data, although the same overall trend is present.

\section{Conclusions} \label{sec:conclusion}
 In this paper, we showed that  a pair-wise $N$-body interaction is equivalent to $N$ one-body interactions with the center of mass of the system. In the attractive and repulsive case, the analytical trajectories are given by harmonic oscillator and exponential solutions, respectively.
 This is in stark contrast to other long-range interactions such Newtonian gravity, where only special case solutions exist for $N > 2$. 
 
It is counter-intuitive that the motion of each particle is independent of the location and motion of all of the other particles of the system, except through the common center of mass location.  The particles all interact with each other, and yet experience no scattering.  This behavior was confirmed with a numerical simulation of the system.


For the attractive  system, the relationships found among the pressure, temperature, and volume are qualitatively similar to what one would expect, pressure and temperature decreasing with increasing volume, and reaching zero for some critical size, as the bound particles have an upper limit on their distance from the center of mass.  A more detailed study comparing the thermodynamics of this system for small containers with that of the ideal gas, or van der Waals gas would be interesting.  For small volumes, the system is expected to behave as free particles in a gas.
 
The mathematical form of the solutions, indicating the independent classical motion of each of the masses, implies  a separable Hamiltonian, which has been investigated in quantum systems of identical particles. \cite{Calogero1969, Calogero1971, BialynickiBirula1985, Lim1971, Vladimiroff1974}  The results presented here show that such a solution is possible in a  quantum equivalent with differing masses.  

The authors are actively working to extend the well-known systems of simple harmonic oscillator pair-wise interacting quantum systems of identical masses \cite{Calogero1971, Lim1971, Vladimiroff1974,Kocik2013,Chen2018,Sowiski2019,ZauskaKotur2000}, to cases where the masses of the particles are not necessarily identical.  In particular, a system with mixed masses seems very promising as an educational tool as a relatively simple model of the strong force and nuclear physics.

\appendix
\section{Collision handling algorithm for particles in a container} \label{sec:collision_algorithm}

 This algorithm uses event-driven collision detection and elastic collision handling to simulate interactions between particles and the walls of a container. This is a discrete collision detection algorithm, \cite{GarciaAlonso1994,Uchiki1983} as opposed to the more computationally expensive continuous collision detection algorithm. \cite{Redon2002}
 
 The algorithm begins by using the analytical trajectories to determine which particles will collide with the container wall in the next time step. This is done by checking if the future position of each particle, \( r(t + \Delta t) \), exceeds the container's radius \( R \). 
For each of these particles, the collision is treated as elastic. The new velocity of the particle is computed on the assumption that it occurs at position $r(t)$ and time $t$. 

The change in momentum for each particle due to the collision is calculated, and the total change in momentum is summed for all particles. This summed momentum change is used to update the velocity of the center of mass (\( V_{\text{cm}} \)). To this point, no particle positions have been changed.

Using the updated center of mass velocity and the current positions and velocities of the particles, new parameters for each particle's trajectories are calculated. The positions and velocities of all particles are then advanced to \( t + \Delta t \), ensuring that no further collisions occur within this time step. The algorithm then repeats this process for subsequent time steps. This collision-handling algorithm has an $\mathcal{O}(\Delta t)$ error in both position and velocity. 

As an example, the data provided in Figs. \ref{fig:swarm_plots} and \ref{fig:Fireball} involves $N =$ 40,000 particles over 500,000 time steps. This simulation runs on a standard laptop in less than 40 seconds. 
For scenarios involving fixed walls, or when expansion into a larger container is permitted, energy conservation is confirmed within computational precision.

\section{Student Projects}\label{sec:projects}
The thermodynamic applications suggest several other investigations for a  student researcher.  Including elastic hard-sphere collisions, as in an ideal gas model, would affect individual trajectories but the state variables of the system would be unaffected.  It might however lead to thermalization on a much shorter time scale.

In this work, particles collide with the a stationary container wall.  Collisions with a moving container would model work done on the system. It would be interesting to explore possible hysteresis effects of a container moving inward and then allowed to expand.

One could also investigate the effects of changing the effective temperature of a fixed container. This could be accomplished by making the collisions with the wall inelastic, simulating a point of impact wall velocity.  The statistical speed distribution of the effective wall velocity would provide the temperature of the wall.

Additionally, these systems can be simulated in situations that include outside forces. 
For example, linear drag could be applied to one or more of the particles, allowing students to illustrate known results.
One could also simulate an $N$-body attractive system released in a uniform gravitational field and allowed to bounce on a hard surface, tracking the motion of the center of mass and the characteristic size of the system as it experiences these outside forces. Selected python programs demonstrating these external interactions have been prepared using glowscript\cite{bartz_trinkets}.
Due to external interactions, these simulations rely on traditional numerical integration, rather than the analytic form of the solutions.   As anticipated, these simulation run times scale as $N^2$, in contrast with the simulations that make use of the analytic solution, which scale linearly with $N$.

\section{Analysis of log-log plots of thermodynamic quantities}
In this supplemental material, we present log-log plots of all thermodynamic quantities from the main text, for both attractive and repulsive cases. A straight line fit on a log-log plot suggests a power-law relationship, with the slope indicating the exponent.

\begin{figure}[htb]
    \centering
    \begin{subfigure}[b]{0.45\textwidth}
        \centering
        \includegraphics[width=\textwidth]{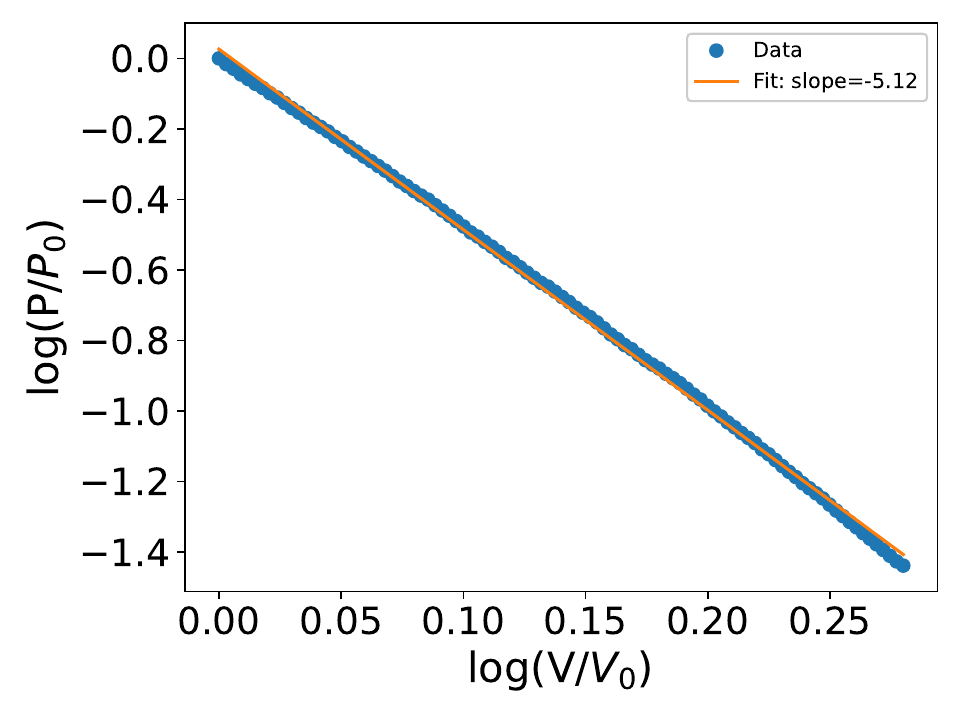}
    \end{subfigure}
    \hfill
    \begin{subfigure}[b]{0.45\textwidth}
        \centering
        \includegraphics[width=\textwidth]{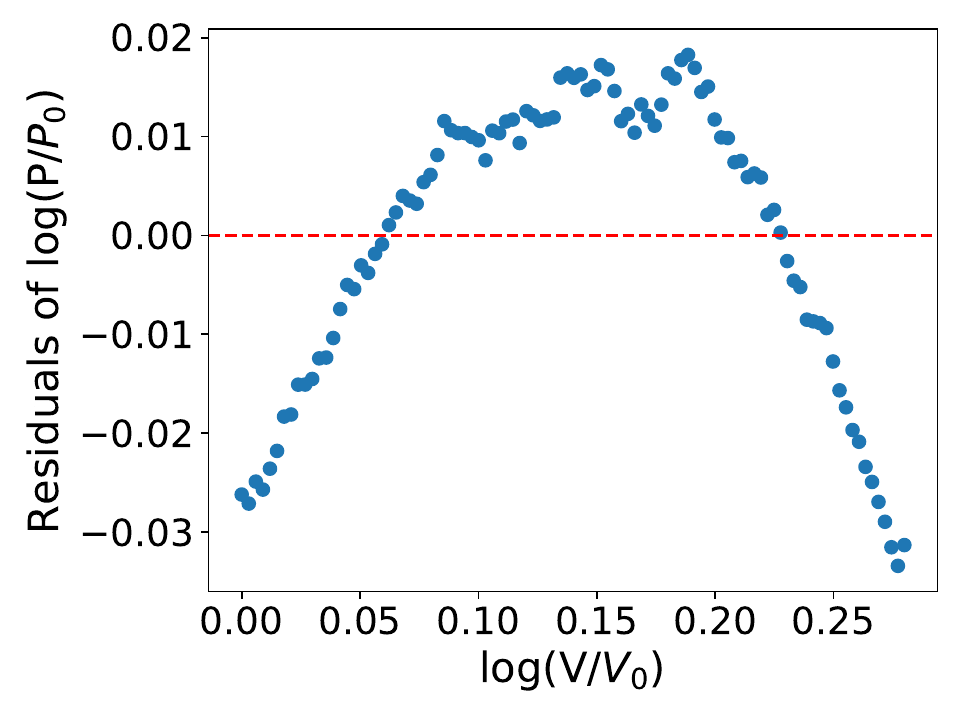}
    \end{subfigure}
    \caption{Attractive case: Log(pressure) vs.~Log(volume) and residuals. The log-log plot suggests a power-law relationship, but the residuals show clear systematic trends, indicating that the linear fit on the log-log plot does not accurately describe the data. \label{fig:attractivePV}}
\end{figure}

\begin{figure}[htb]
    \centering
    \begin{subfigure}[b]{0.45\textwidth}
        \centering
        \includegraphics[width=\textwidth]{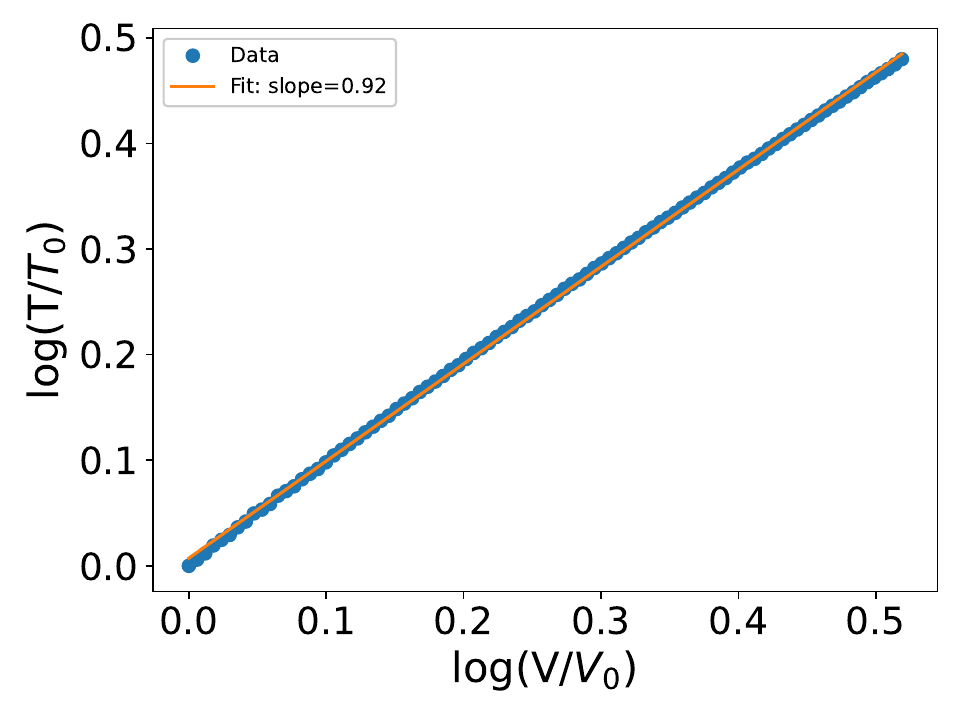}
    \end{subfigure}
    \hfill
    \begin{subfigure}[b]{0.45\textwidth}
        \centering
        \includegraphics[width=\textwidth]{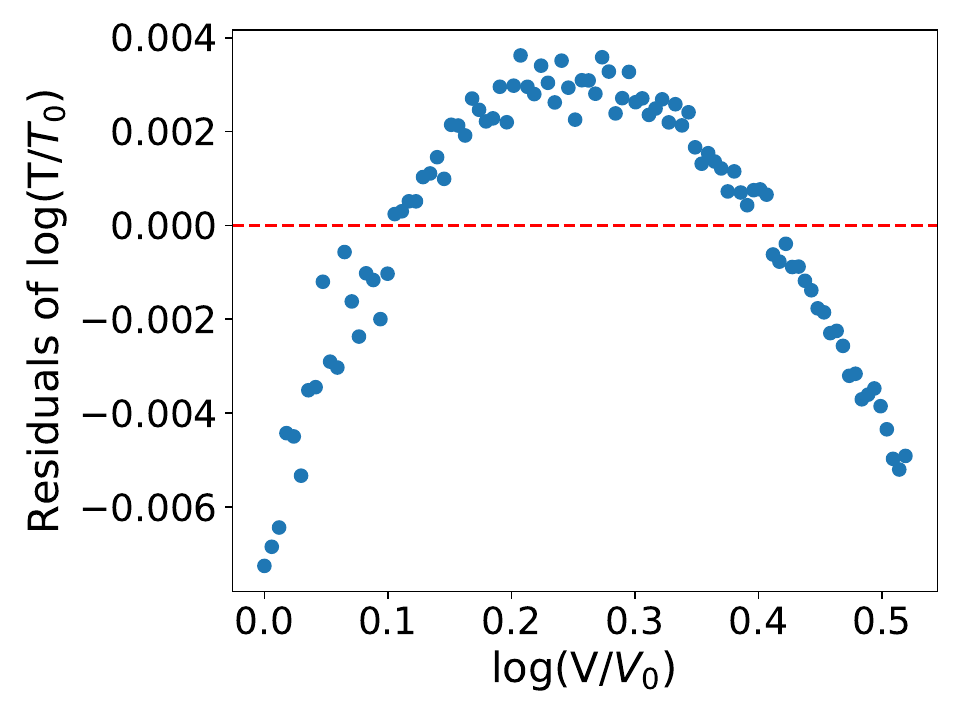}
    \end{subfigure}
    \caption{Repulsive case: Log(temperature) vs.~Log(volume) and residuals. The log-log plot indicates an apparent power-law relationship, but the systematic trend in the residuals undermines the validity of the fit. \label{fig:repulsiveTV}}
\end{figure}

While the best-fit lines appear visually convincing, residual analysis reveals significant issues. In key cases, residuals exhibit systematic behavior, contradicting the assumption of a linear log-log fit. This is evident in the attractive case for pressure vs.volume (Fig.\ref{fig:attractivePV}) and the repulsive case for temperature vs.volume (Fig.\ref{fig:repulsiveTV}).

\begin{figure}[htb]
    \centering
    \begin{subfigure}[b]{0.45\textwidth}
        \centering
        \includegraphics[width=\textwidth]{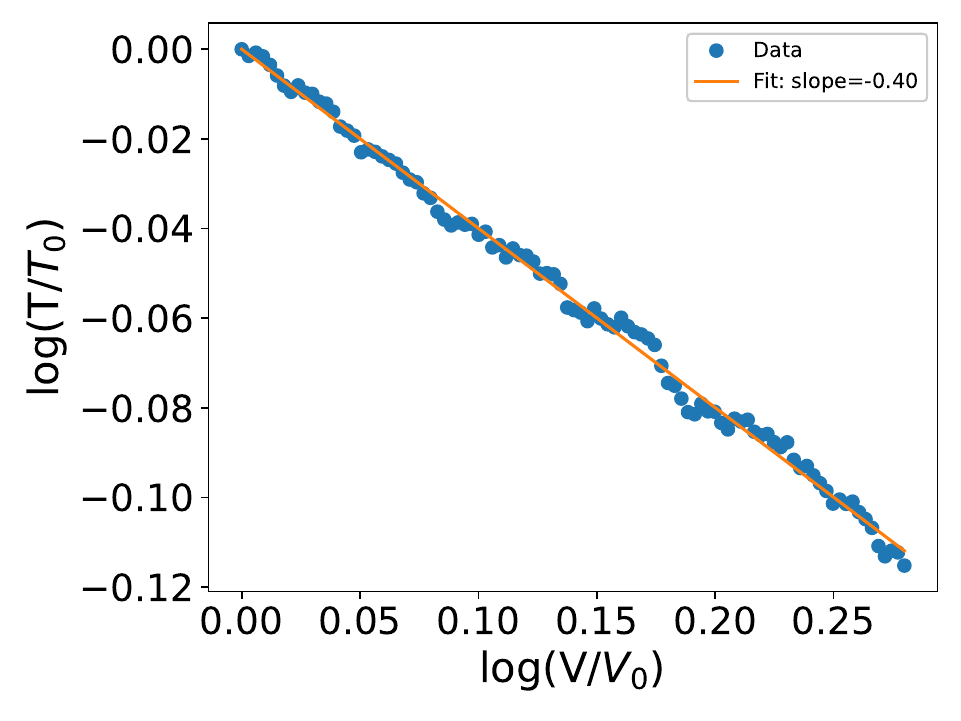}
    \end{subfigure}
    \hfill
    \begin{subfigure}[b]{0.45\textwidth}
        \centering
        \includegraphics[width=\textwidth]{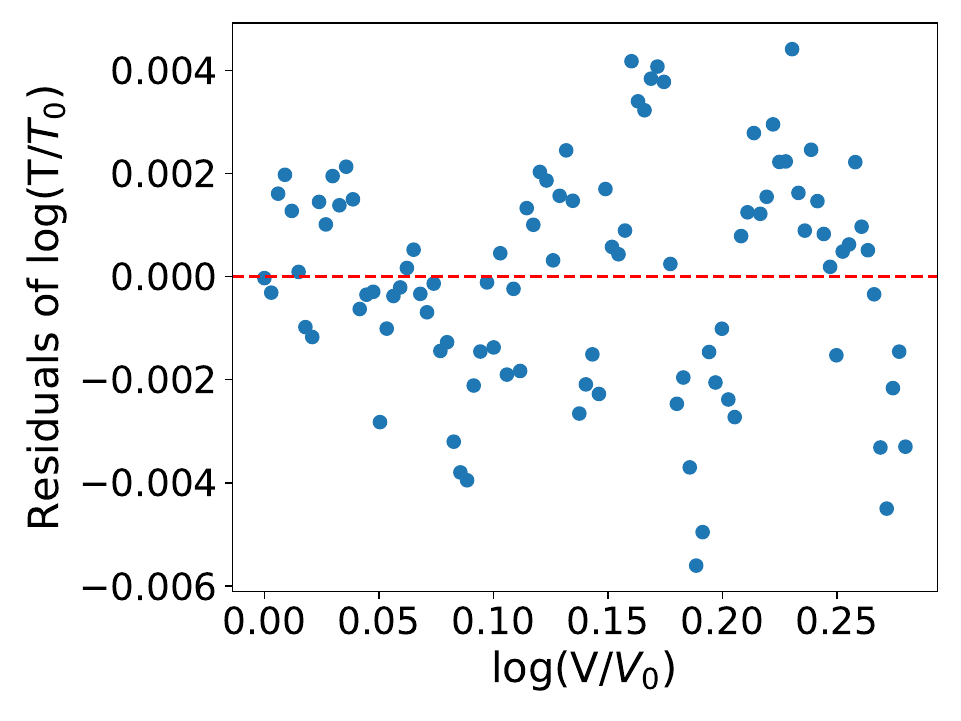}
    \end{subfigure}
    \caption{Attractive case: Log(temperature) vs.~Log(volume) and residuals. The log-log plot shows temperature as a function of volume, with the best-fit line indicating an apparent power-law relationship. However, the residuals exhibit a wave-like pattern, indicating systematic deviations and suggesting that the power-law fit is not adequate. \label{fig:swarm_temp_vol_loglog}}
\end{figure}

\begin{figure}[htb]
    \centering
    \begin{subfigure}[b]{0.45\textwidth}
        \centering
        \includegraphics[width=\textwidth]{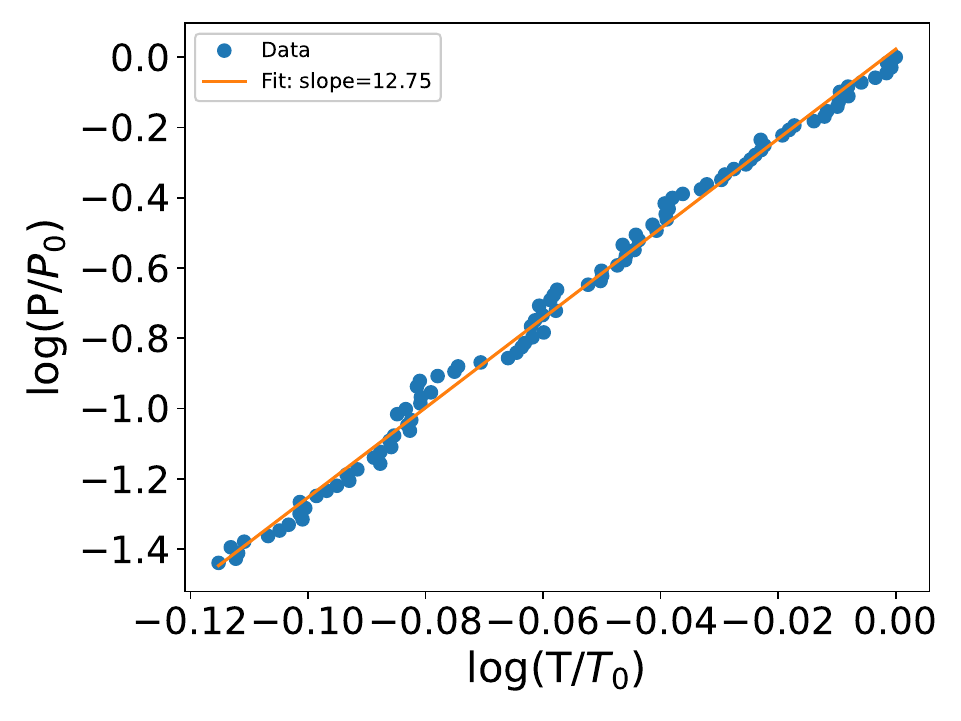}
    \end{subfigure}
    \hfill
    \begin{subfigure}[b]{0.45\textwidth}
        \centering
        \includegraphics[width=\textwidth]{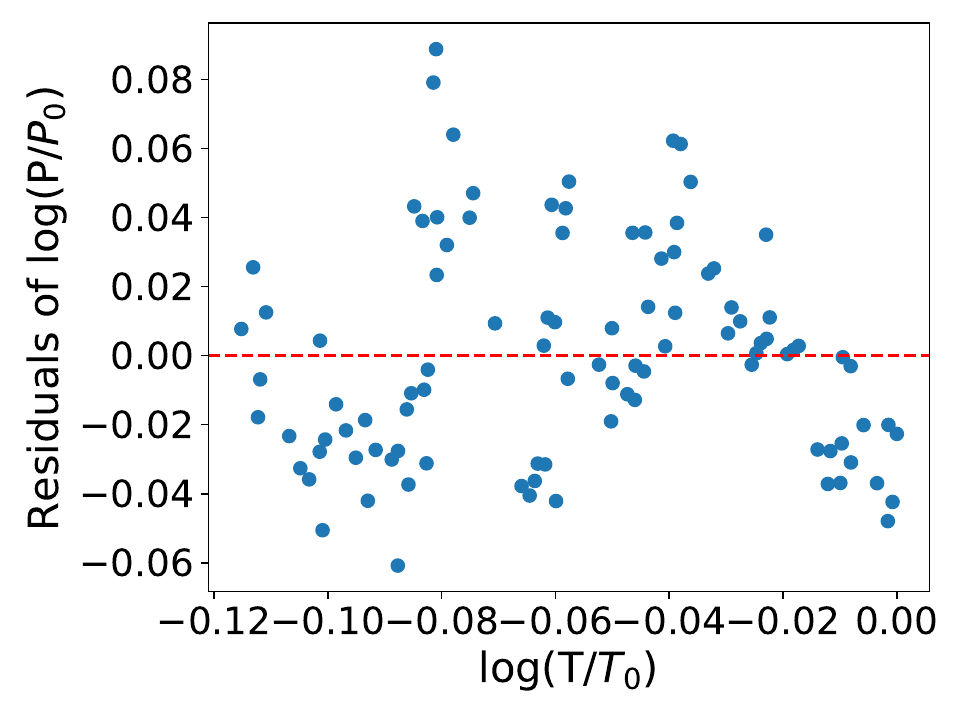}
    \end{subfigure}
    \caption{Attractive case: Log(pressure) vs.~Log(temperature) and residuals. However, the residuals exhibit a wave-like pattern, indicating systematic deviations and suggesting that the power-law fit is not adequate. \label{fig:swarm_p_temp_loglog}}
\end{figure}

In the attractive case, temperature data fluctuations dominate the residuals (Figs.~\ref{fig:swarm_temp_vol_loglog}, \ref{fig:swarm_p_temp_loglog}). These fluctuations are temporally correlated, as seen in the wave-like residual patterns, reflecting non-random, systematic deviations. Conversely, pressure data shows less fluctuation and more systematic residual behavior.

\begin{figure}[htb]
    \centering
    \begin{subfigure}[b]{0.45\textwidth}
        \centering
        \includegraphics[width=\textwidth]{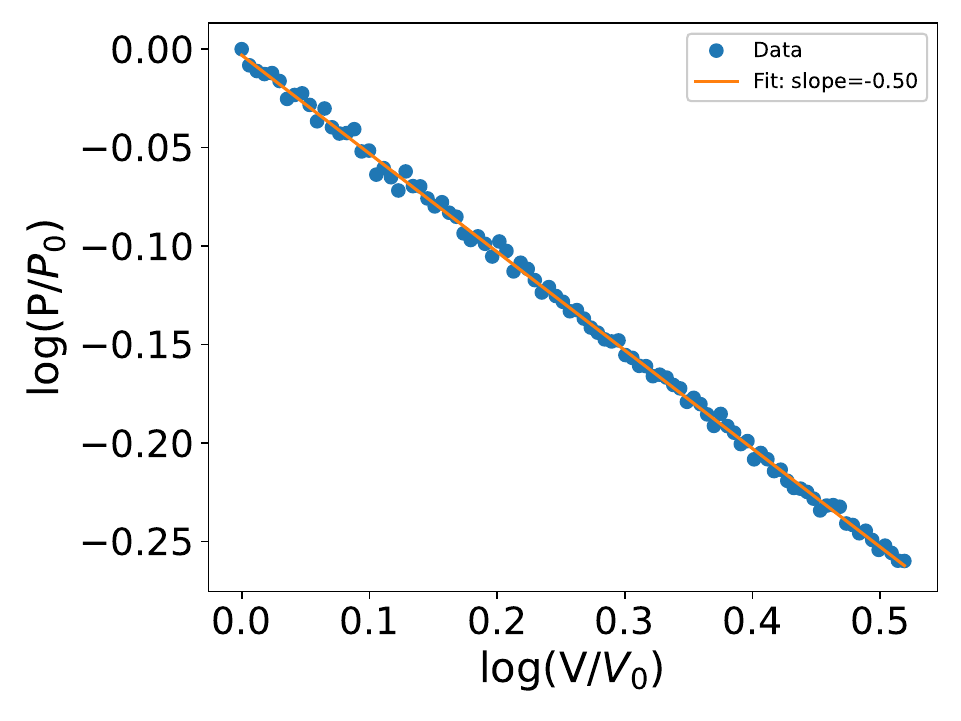}
    \end{subfigure}
    \hfill
    \begin{subfigure}[b]{0.45\textwidth}
        \centering
        \includegraphics[width=\textwidth]{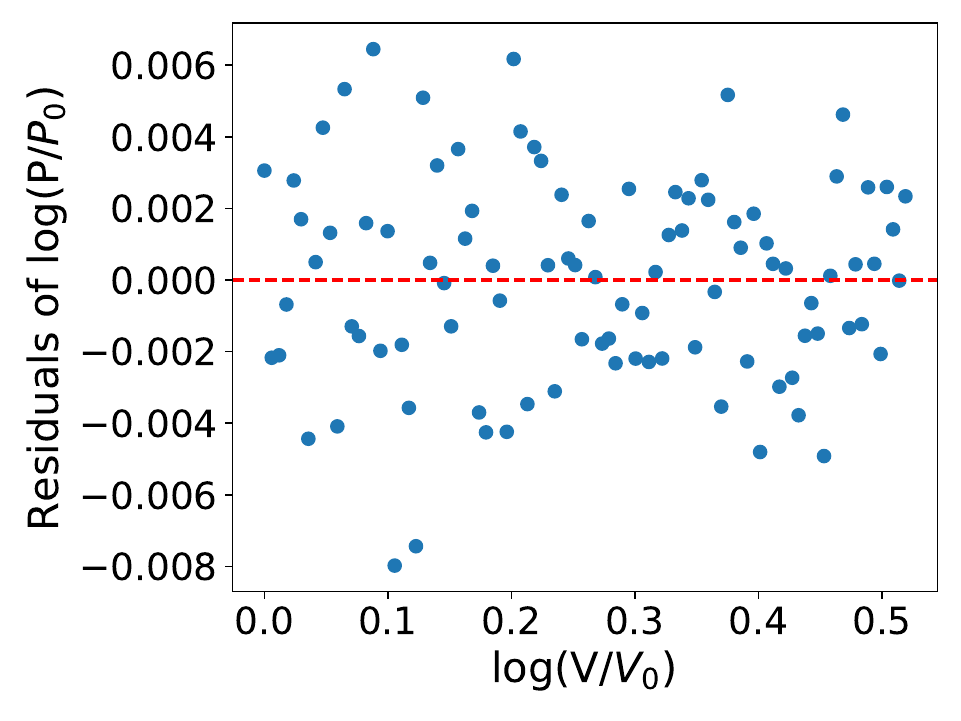}
    \end{subfigure}
    \caption{Repulsive case: Log(pressure) vs.~Log(volume) and residuals. \label{fig:repulsivePV}}
\end{figure}

\begin{figure}[htb]
    \centering
    \begin{subfigure}[b]{0.45\textwidth}
        \centering
        \includegraphics[width=\textwidth]{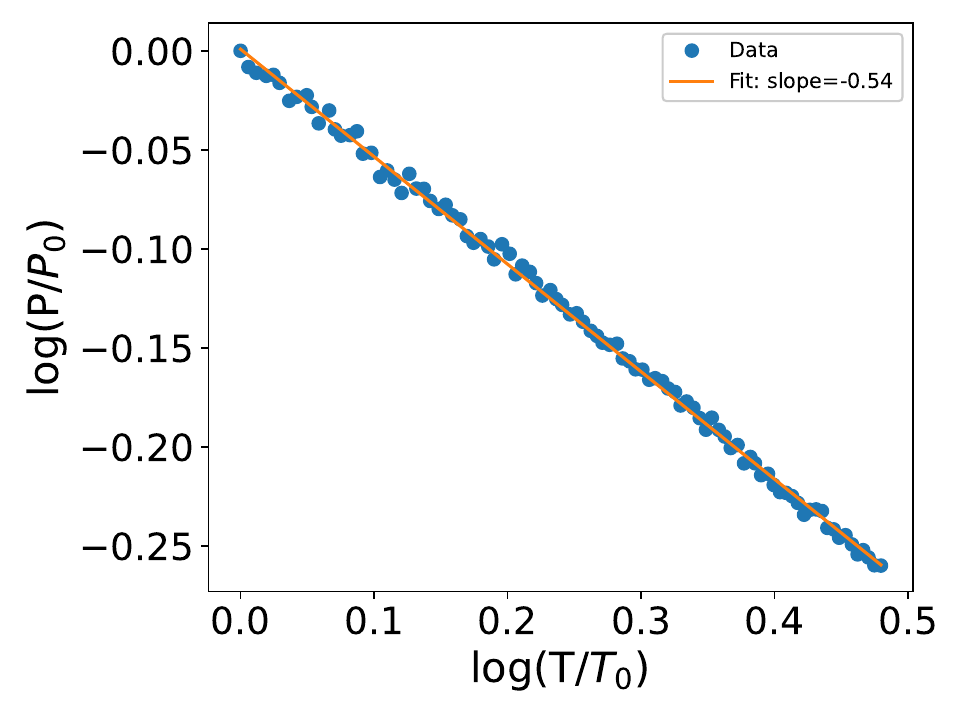}
    \end{subfigure}
    \hfill
    \begin{subfigure}[b]{0.45\textwidth}
        \centering
        \includegraphics[width=\textwidth]{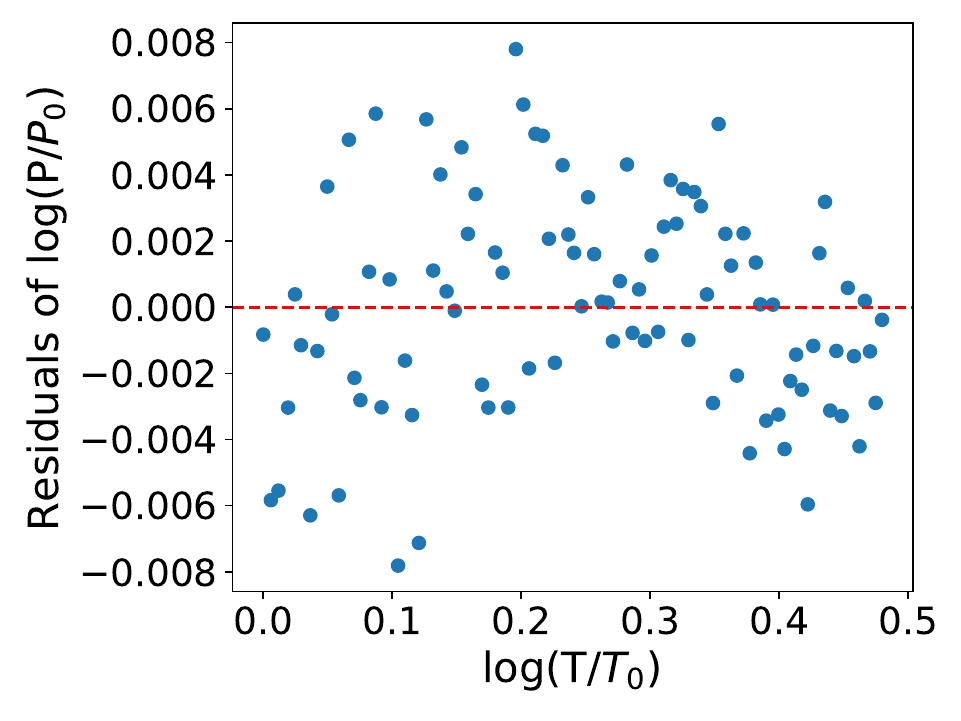}
    \end{subfigure}
    \caption{Repulsive case: Log(pressure) vs.~Log(temperature) and residuals. \label{fig:repulsivePT}}
\end{figure}

In the repulsive case, pressure data shows greater fluctuation than temperature (Figs.~\ref{fig:repulsivePV}, \ref{fig:repulsivePT}). While the pressure is autocorrelated in time, residuals do not exhibit clear systematic deviation or patterns. This inconsistency underscores why the relationships among thermodynamic variables require further analysis.

\bibliography{biblio.bib}
\bibliographystyle{utphys}
 
\end{document}